\documentclass{iau} 
\usepackage{graphicx}
\usepackage{natbib}
\usepackage{amsmath}
\usepackage{multirow}
\usepackage[svgnames]{xcolor}
\usepackage{url}

\def\aap{A\&A}
\def\aaps{A\&AS}
\def\apj{ApJ}
\def\araa{ARA\&A}
\def\mnras{MNRAS}

\title[Massive stars at low $Z$] 
{Massive star evolution: \\ feedbacks in low-$Z$ environment}

\author[Ekstr\"om et al.]   
{Sylvia Ekstr\"om$^1$
 \and Georges Meynet $^1$ \and Cyril Georgy$^1$ \and José Groh$^2$ \and Arthur Choplin $^1$ \and Hanfeng Song$^3$}

\affiliation{$^1$Department of Astronomy, University of Geneva, \\ Maillettes 51,
CH-1290, Versoix, Switzerland \\ email: {\tt sylvia.ekstrom@unige.ch} \\[\affilskip]
$^2$Trinity College, University of Dublin, \\ Dublin 2, Republic of Ireland \\[\affilskip]
$^3$College of Physics, Guizhou University, \\Guiyang City, 550025 Guizhou Province, PR China}

\pubyear{2019}
\volume{344}  
\jname{Dwarf galaxies: from the deep Universe to the present}
\editors{S. Stierwalt \& K. McQuinn, eds.}
\begin{document}

\maketitle

\begin{abstract}
Massive stars are the drivers of the chemical evolution of dwarf galaxies. We review here the basics of massive star evolution and the specificities of stellar evolution in low-$Z$ environment. We discuss nucleosynthetic aspects and what observations could constrain our view on the first generations of stars.
\end{abstract}
\firstsection 
\section{Stellar evolution}
\subsection{The basics}
\begin{figure}[t]
\begin{center}
 \includegraphics[width=8.5cm]{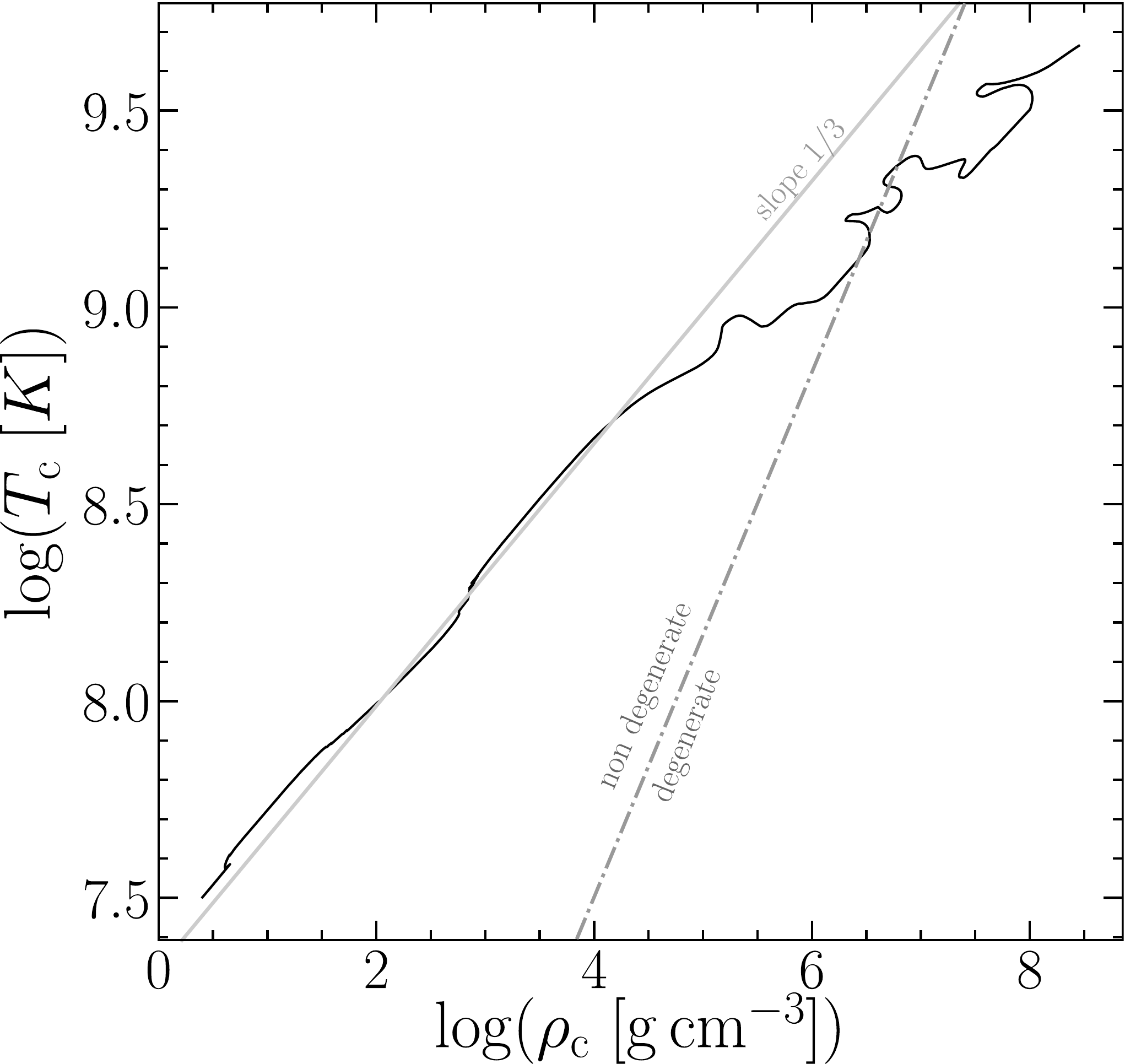} 
 \caption{Path of a $25\,M_\odot$ model in the central $T$ vs $\rho$ diagram. The slope 1/3 of perfect gas is shown as a solid grey line. The dash-dotted grey line shows the 2/3 slope of the limit between non-degenerate and non-relativistic degenerate gas.}
   \label{rhoT25}
\end{center}
\end{figure}

Stars are gaseous spheres that can be considered as in hydrostatic and thermal equilibrium during most of their lifetime. Four basic equations are used to describe the structure:
$$\begin{array}{lcll}
\frac{\text{d}M_r}{\text{d}r} & = & 4\pi\,r^2\,\rho & \text{: mass conservation} \\[.2cm]
\frac{\text{d}P}{\text{d}r} & = & -\frac{GM_r}{r^2}\rho & \text{: hydrostatic equilibrium} \\[.2cm]
\frac{\text{d}L_r}{\text{d}r} & = & 4\pi\,r^2\,\rho\,\left(\varepsilon+\varepsilon_{grav}\right) & \text{: energy conservation}\\[.2cm]
\frac{\text{d}T}{\text{d}r} & = & -\frac{3\kappa\rho}{4acT^3}\frac{L_r}{4\pi\,r^2} & \text{: radiative transfer}
\end{array}$$
In most models, $\epsilon$ and $\kappa$ are interpolated from tables (NACRE\footnote{\url{http://pntpm.ulb.ac.be/Nacre/nacre.htm}}, OPAL\footnote{\url{https://opalopacity.llnl.gov}}, OP project\footnote{\url{http://opacities.osc.edu}}, among others). In order to close the set of equations, we need a fifth one giving the relation between the pressure $P$, the density $\rho$, and the temperature $T$. This is done using an equation of state (EOS) usually of the form $\Delta \ln \rho = \alpha \Delta \ln P - \delta \Delta \ln T$. Through the first equation, we get a relation between $r$ and $\rho$: $\frac{\Delta \rho}{\rho} = -3 \frac{\Delta r}{r}$, and through the second one, a relation between $r$ and $P$: $\frac{\Delta P}{P} = -4 \frac{\Delta r}{r}$, which yields a relation between $P$ and $\rho$: $\Delta\ln P=\frac{4}{3}\Delta\ln\rho$. Now if we use our favourite EOS, we can determine the relation between $T$ and $\rho$:
$$\Delta \ln T = \left( \frac{4\alpha-3}{3\delta} \right) \Delta\ln\rho.$$
All the physics of the matter is contained in the parameter $\alpha$ and $\delta$. For a perfect gas (PG), we have $\alpha=\delta=1$ since $P=\frac{k}{\mu m_H}\rho T$. Hence the slope in a $T$ vs $\rho$ diagram is 1/3. The cores of stars during the first phases of their evolution can be approximated by a perfect gas. As we can see in Fig.~\ref{rhoT25}, they indeed follow a path of slope 1/3 during the H- and He burning phase. When compressed too much, the perfect gas can become degenerate.
To find the limit between these two states of the matter, we look for the region where both pressures are equal:
$$P_\text{PG}=\frac{k}{\mu m_H}\rho T= K_1\left(\frac{\rho}{\mu_e}\right)^{5/3} = P_\text{deg}$$
which yields a slope of 2/3. This limit being steeper than the evolution one, the stars reach the limit at one point or the other of their evolution. The phase at which they reach it determines the type of stars they are. We can roughly summarise the types as follows:
\begin{description}
  \item[low-mass stars: ] they reach the limit right after H-burning;
  \item[intermediate-mass stars: ] they enter the degenerate zone after He-burning;
  \item[massive stars: ] they are able to follow all the burning phases up to Si-burning.
\end{description}

\subsection{Massive star evolution}
The detailed evolution can be followed in the $T_\text{c}$ vs $\rho_\text{c}$ diagram, but this is accessible only to models. The evolution of real stars can be observed only in the Hertzsprung-Russell (HR) diagram. Unfortunately, during the central C-burning phase, a decoupling between the core and the envelope occurs, because their characteristic timescales become different by orders of magnitude. But while the final stages of a star are hidden behind the veil of an unchanging envelope, the excursion it makes in the HRD during the two first stages can be characterised. \citet{Conti1975a} proposed a scenario of filiation between the observed types of single massive stars, of which an updated version is shown on Table~\ref{Conti_scenario}.
\begin{table}[h]
  \begin{center}
  \caption{Modified Conti scenario \citep{Conti1975a} for the filiation of massive stars as they can be observed.}
  \label{Conti_scenario}
 {\scriptsize
\begin{tabular}{|lllc|}
\hline
$M > 60\, M_\odot$: & {\small O $\rightarrow$ Of/WNL $\rightarrow$ LBV $\rightarrow$ WNL $\rightarrow$ WC $\rightarrow$ WO} & $\rightarrow$ SNIbc? & \multirow{4}{*}{\large {\bf WR}}\\
 & & & \\
{$M = 40-60\, M_\odot$}: & {\small O $\rightarrow$ BSG $\rightarrow$ LBV $\rightarrow$ WNL $\rightarrow$ (WNE) $\rightarrow$ WC} & $\rightarrow$ SNIbc? & \\
                                        & \hfill {\small $\rightarrow$ WC $\rightarrow$ WO} &  $\rightarrow$ SNIbc? & \\
\hline
\multicolumn{1}{l}{\null} & & & \multicolumn{1}{c}{\null} \\
\multicolumn{1}{l}{$M = 30-40\, M_\odot$:} & {\small O $\rightarrow$ BSG $\rightarrow$ RSG $\rightarrow$ WNE$\rightarrow$ WCE} & $\rightarrow$ SNIbc & \multicolumn{1}{c}{\null} \\
\multicolumn{1}{l}{\null} & & & \multicolumn{1}{c}{\null} \\
\hline
{$M = 25-30\, M_\odot$}: & {\small O $\rightarrow$ (BSG) $\rightarrow$ RSG $\rightarrow$ (YSG? LBV?)} & $\rightarrow$ SNII-L/b & \multirow{3}{*}{\large {\bf RSG}}\\
 & & & \\
{$M = 10-25\, M_\odot$}: & {\small O/B $\rightarrow$ RSG $\rightarrow$ (Ceph. loop for $M < 15\, M_\odot$) $\rightarrow$ RSG} & $\rightarrow$ SNII-P & \\
\hline
\end{tabular}
}
\end{center}
\end{table}
We can roughly divide the massive stars in two categories: the stars that end up as red supergiants (RSG), and the ones that end up as Wolf-Rayet (WR) stars. The real filiation depends on many factors, as the mass loss experienced, the rotation rate, the presence or not of a magnetic field, the multiplicity status, and of course the metallicity. In the next section we will review a few of them.
\subsection{Beyond the basics}
Let us first turn towards the {\bf stellar winds}. They play a dominant role in the evolution of massive stars \citep{Langer2012a}, being determinant for the endpoint location in the HR diagram \citep{Groh2013c}. Unfortunately, the precise mechanism of mass loss is not at reach in 1D simulations, where we have to apply prescriptions given in the literature \citep[][etc...]{Reimers1975a,deJager1988a,Kudritzki1987a,Kudritzki2000a,Nugis2000a,Vink2000a,Vink2001a,vanLoon2005a,Grafener2007a}. Some of these prescriptions are empirical or semi-empirical, others are theoretical. They often cover only a narrow validity domain, so models have to switch from one to another. Some of them include the clumpiness of the wind, others don't, and it is not yet clear by which amount the rates have to be reduced because of this clumpiness. Some stars might go through bursts of mass loss (like LBVs), while in stellar modelling we are bound to use averaged rates. The mass-loss rates, even applied for a very short time, have a huge impact on the evolutionary track of the star \citep{Groh2019a}. Hence a comparison between massive stars and their models is rather a check for the mass-loss recipe used in the model than anything else.

\begin{figure}[t]
\begin{center}
 \includegraphics[width=5cm]{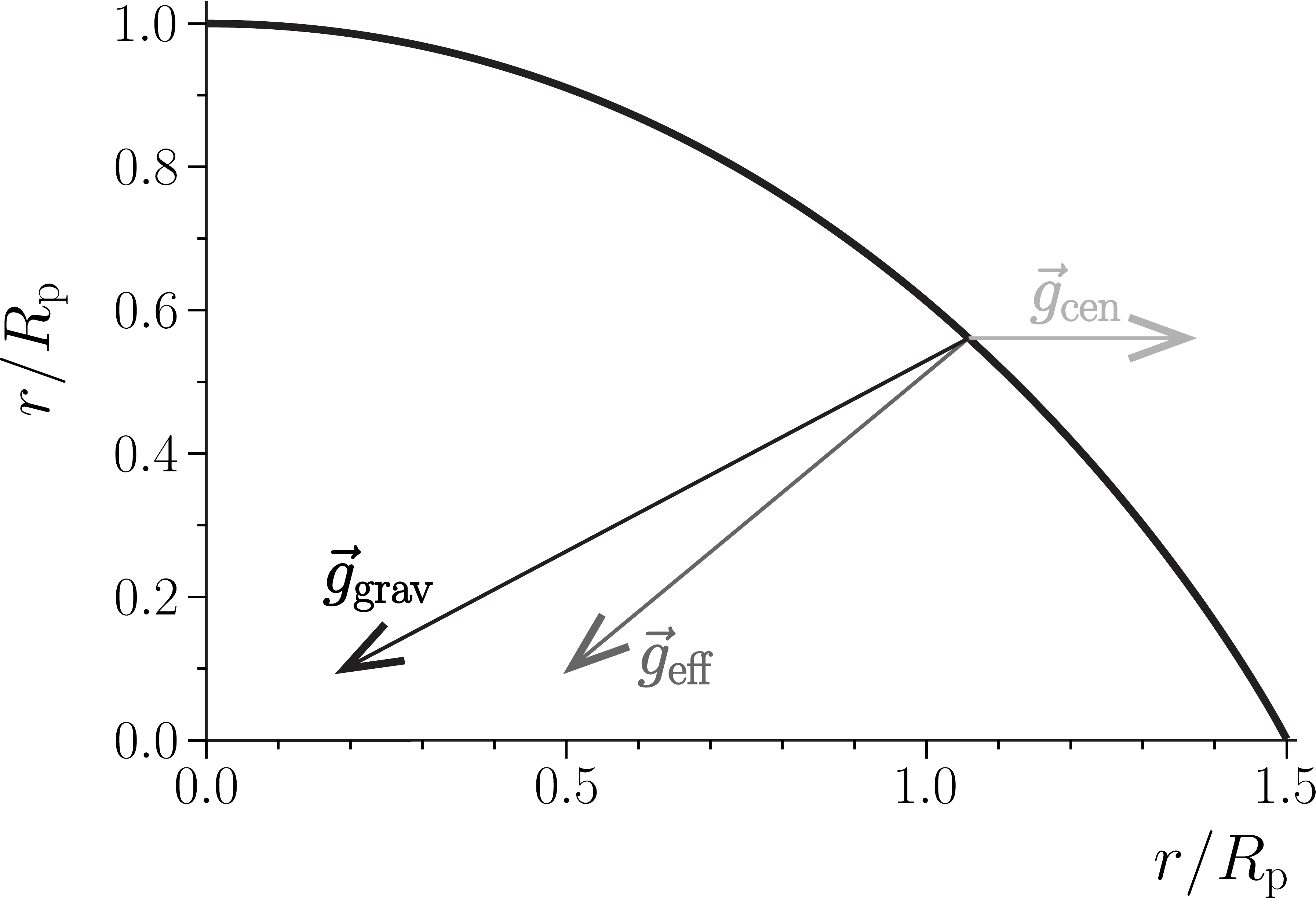} \hspace{0.5cm} \includegraphics[width=5cm]{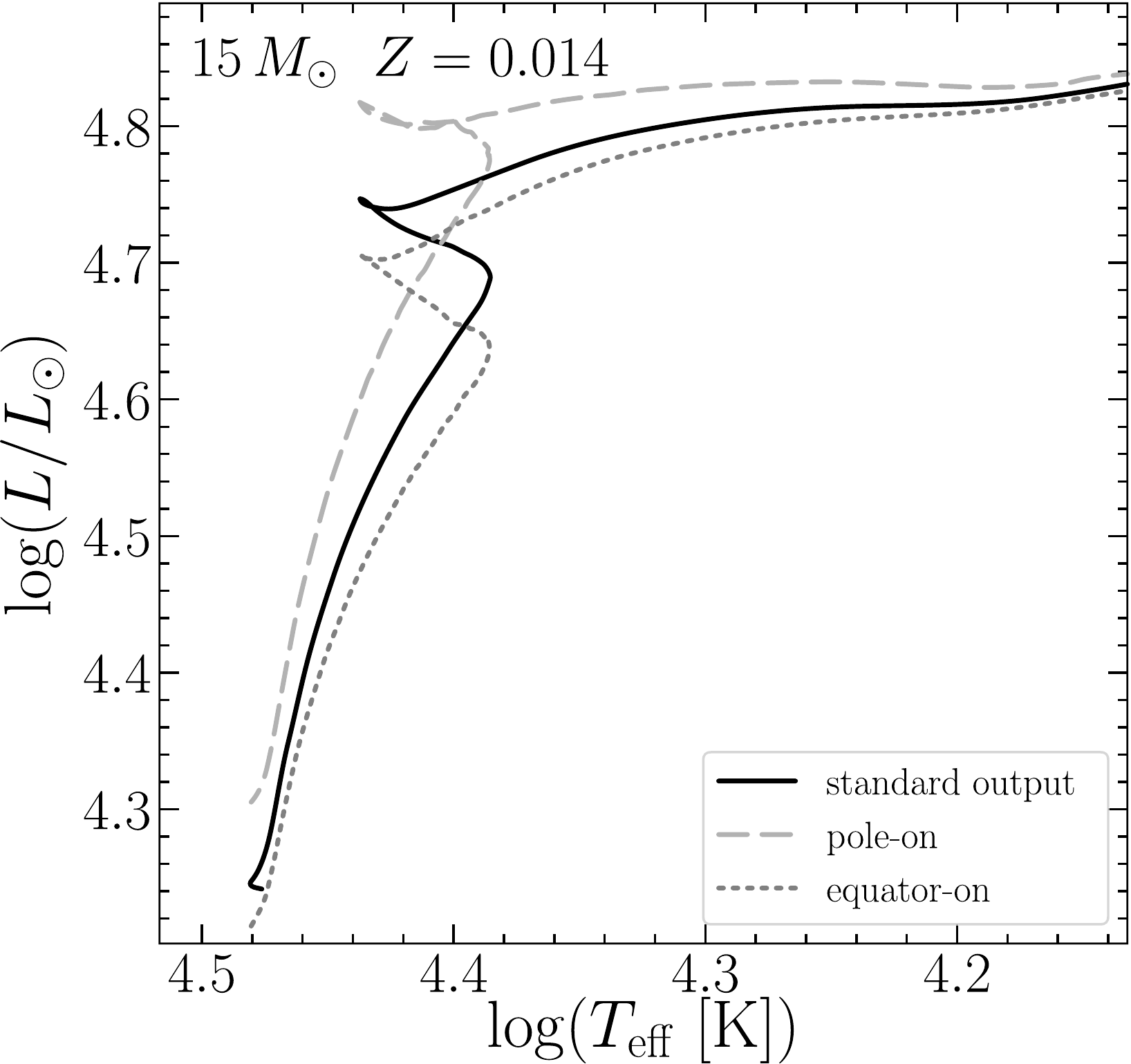}
 \caption{{\it Left:} shape of a critically rotating star, and the effective gravity $g_\text{eff}$. Figure from \citet{Georgy2010a}. {\it Right:} HR diagram of a $15\,M_\odot$ model at $Z=0.014$, with an initial rotation rate $\Omega/\Omega_\text{crit}=0.80$ (black solid line). Grey lines: $T_\text{eff}$ and $L$ deduced from the isotropic flux hypothesis when the same star is observed pole-on (dashed) or equator-on (dotted).}
   \label{geff}
\end{center}
\end{figure}

Another physical ingredient modifies drastically the evolution of stars: {\bf rotation}. Rotating stars are expected to present a modified gravity, because rotation induces an oblateness of the star (see Fig.~\ref{geff} {\it left}).
The stellar characteristics become dependent on the co-latitude considered (see Fig.~\ref{geff} {\it right}), because there is a relation between gravity and effective temperature \citep{vonZeipel1924a}:
$$T_\text{eff} = T_\text{eff}(\Omega,\theta) = \left[\frac{L}{4\pi\sigma\,G\,M^\star}\ g_\text{eff}(\Omega,\theta) \right]^{1/4}.$$
Note that a more recent $T_\text{eff}-g_\text{eff}$ relation has been proposed by \citet{EspinosaLara2011a}, but the qualitative result remains the same. The oblateness however becomes significative only for rapid rotation rates ($\Omega/\Omega_\text{crit}>0.70$).

Rotation has an impact on stellar winds \citep{Owocki1997a,Petrenz2000a}, increasing them by a factor of
$$
\frac{\dot{M}(\Omega)}{\dot{M}(0)} = \left[\frac{\left( 1-\Gamma_\text{Edd} \right)}
{\left( 1-\frac{\Omega^2}{2\pi\,G\,\rho_m} - \Gamma_\text{Edd} \right)}\right]^{\frac{1}{\alpha}-1}
$$
\citep[][see however \citealt{Muller2014a}]{Maeder2000a}. The mass flux is expected to be anisotropic, with a difference between polar and equatorial mass loss up to more than a factor of 3 for very rapid rotation rates \citep{Georgy2011a}.

\begin{figure}[t]
\begin{center}
 \includegraphics[width=8.5cm]{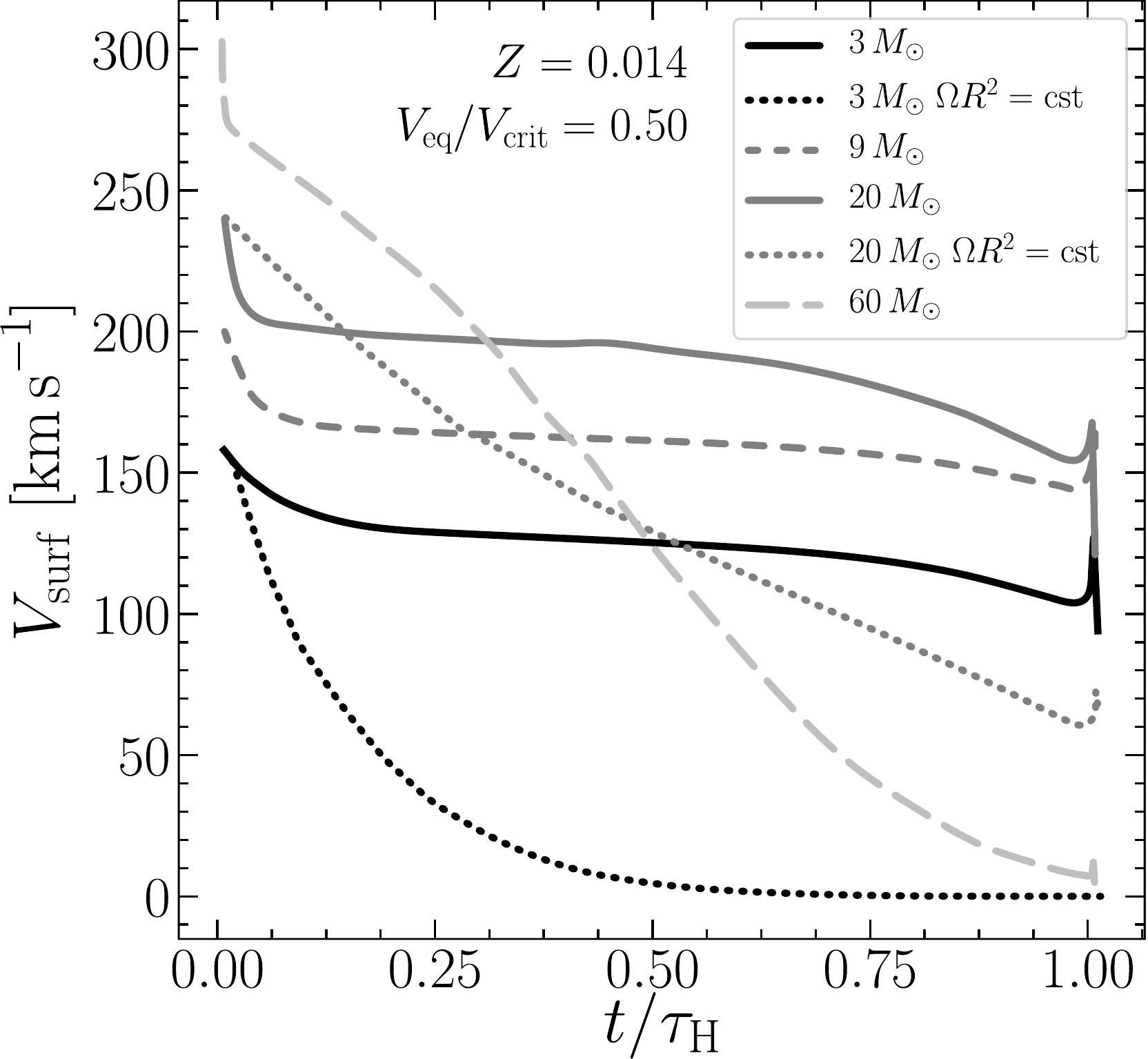}
 \caption{Evolution of the equatorial velocity during the main sequence lifetime for stellar models of $3$, $9$, $20$, and $60\,M_\odot$ at solar metallicity with an initial rotation rate of $V_\text{eq,ini}/V_\text{crit}=0.50$. Another $3$ and $20\,M_\odot$ models have been computed with no transport mechanism in the interior, only the conservation of angular momentum (dotted lines).}
   \label{Vevol}
\end{center}
\end{figure}
Another effect of rotation is to add a mixing process inside the star, mainly through two mechanisms. First, the thermal imbalance inside the star generates a large-scale current called the meridional circulation. Second, a shear, due to the differentially-rotating layers, is generated in the radiative zones. The internal mixing provides a coupling between the contracting core and the expanding envelope. While under the sole action of angular-momentum conservation, the surface velocity would decrease rapidly (Fig.~\ref{Vevol}, dotted lines), the angular momentum brought from the accelerating core maintains it more or less constant during the MS (Fig.~\ref{Vevol}, solid lines).
The winds also have an impact on the rotational velocity of the star, since by removing mass they remove also angular momentum. If the anisotropy is strong, there is less angular momentum removed than in the spherical case, however massive stars with strong radiative mass loss are expected to have their surface efficiently braked. Actually, the net evolution of the surface velocity is the subtle result of both the braking and the transport processes inside the star.

The mixing transports not only angular momentum, but also the chemical species. Rotating stars are expected to present a modification of their surface composition during the evolution, even before any dredge-up occurs. The mixing is stronger in higher mass stars and in more rapid rotators.

Massive stars often live in {\bf binary} systems \citep{Sana2012a} and many apparently single stars might be indeed undetected binaries \citep{deMink2014a}. Binarity influences all the aspects described above. Close binaries undergo a tidal mixing that modifies their chemical composition and their surface velocity, driving them to get synchronised with respect to the orbital angular velocity. When a mass-transfer episode occurs, they can be the mass donor or the mass gainer, and lose or accrete angular momentum as well as mass. All these aspects have a huge impact on the endpoint of the evolution. Modelling one star is already a difficult task, but modelling binaries is even worst. The parameter space to explore in population synthesis is huge, since for the basic combination of mass, metallicity and rotation rate the full parameter space of binarity has to be added: mass ratio and separation. Different approaches are usually followed. For some specific cases, a full computation of the two components with a more or less detailed binary physics can be undertaken. For the comparison with population of real stars, either crude binary models with very simplified physics are computed, or a population synthesis code is used, based on single star models for which binarity prescriptions are applied, like period and mass ratio distributions, efficiency for the mass transfer, or the tidal influence on rotation.

\section{\boldmath Low-$Z$ environments}
\subsection{$Z$ effects on stellar evolution}
A low content in metals has two main effects on massive stars. First the opacity is lower so the star is more compact. Second, since massive stars rely mainly on the CNO cycle to sustain their gravity, the low content in carbon obliges them to contract longer to reach a higher central temperature able to get the nuclear energy production at a sufficient level. Low-metallicity stars are thus hotter and more luminous than their metallic counterparts, and their radius is smaller.

The deficiency in metals has also an impact on the strength of the radiative winds, which are expected to be lower. In the single star frame, it is thus more difficult to form WR stars at low metallicity \citep{Georgy2013b,Groh2018a}.

\subsection{$Z$ effects on rotation}
The greater compactness of the star makes the meridional circulation weaker, which in turn increases the shear, because the gradient of $\Omega$ is steeper inside the star. Also the diffusion time is shorter since $t_\text{diff}\propto \frac{R^2}{D}$, with $D$ larger and $R$ smaller. The relative surface enrichment is thus expected to be quicker and stronger. More massive stars are more efficiently mixed than less massive ones. The mixing that can be expected at the surface of a star is thus a function not only of the surface velocity, but also of the mass, the metallicity, the evolutionary phase, the possible binarity, or the presence or not of a magnetic field \citep{Maeder2009b}.

When rotation is rapid, it might be a solution to produce WR stars at low $Z$, through the scenario of chemically homogeneous evolution  \citep[CHE, see][]{Szecsi2015a}. Of course, the rotation rate at which CHE occurs depends a lot on the physics considered in the models, so precise values cannot be given and this scenario for forming WR stars must be studied further.
\subsection{$Z$ effects on nucleosynthesis}
Because of the compactness of low-$Z$ stars, different zones of combustion can be in contact through mixing: some carbon produced in the He-burning zone (core or shell) can be mixed out toward the H-burning shell and boost the CNO cycle in it, producing primary nitrogen \citep{Meynet2002a,Hirschi2007a,Yoon2012a}. This nitrogen can be mixed backwards toward the He-burning region, producing $^{22}$Ne, which is a seed for $s$-process elements through the $n$-producing reaction $^{22}$Ne($\alpha,n$) \citep{Frischknecht2012a,Frischknecht2016a,Choplin2016a}.
\subsection{$Z$ effects on binarity}
It is not yet clear wether the binary fraction is metallicity-dependent. However, the effects of the binarity are modified by $Z$. Since the stars are more compact, they undergo mass-transfer episodes later in the evolution, or even may avoid it \citep{Song2016a,Gotberg2018a}. This has strong consequences on the expected outcome of binaries.
\section{Chemical enrichment by massive stars}
Because of lifetime considerations, massive stars are the first actors in the early chemical enrichment of galaxies. For the same considerations, no direct observations are possible, these stars being long gone. We rely on indirect observations to constrain the stellar models and try and understand the very first stellar generations.
\subsection{Winds and/or supernova?}
We saw that low-$Z$ stars are compact, lose little mass through radiatively-driven winds and end their life with a large core. This means that the stellar matter is strongly bound at the end of the evolution, which might prevent them to explode, or let them do so only in a faint supernova \citep{OConnor2011a,Ugliano2012a,Sukhbold2014a,Sukhbold2016a,Pejcha2015a,Ertl2016a,Muller2016a,Ebinger2018a}. In that case, their contribution to the enrichment of the surrounding medium would be only or mostly through winds. Another possibility is the opposite: since they retain almost all the angular-momentum content they had at birth, they might explose in a very energetic magneto-rotational explosion, with the emission of long soft gamma-ray bursts. In that case, it is not clear whether the host galaxy is able to retain the very fast ejected matter if it is a dwarf galaxy. Also the ejecta might be highly asymmetrical in case of a jet-powered supernova \citep{Papish2015a}, leaving a large part of the surrounding medium to be enriched only by the winds.
\subsection{Chemical imprints of early stellar generations}
Chemical evolution models for the Galaxy show that in order to reproduce the N/O and C/O ratios as a function of O/H in the solar vicinity, rapidly-rotating low-$Z$ models have to be included for their ability to synthesise primary nitrogen \citep{Chiappini2006a,Pettini2008a}.

To probe the very first generations of stars, we have to observe the most metal-poor stars in the halo of our Galaxy. They are expected to be born from a matter enriched by only one or very few pre-existing massive stars. A striking feature of the extremely metal-poor low-mass stars is that below [Fe/H] $=-3$, the population is dominated by the so called CEMP-no stars: carbon-enriched extremely metal-poor stars without $r$- or $s$-process elements. Most of them present not only an enhanced C abundance, but also excesses in N and O. \citet{Maeder2015b} have shown that the variety of abundances found at the surface of such stars can be naturally explained by various degrees of back and forth mixing between the H- and He-burning zones followed by matter ejection. As low-$Z$ stars are supposed to lose very little mass through radiatively-driven winds, the matter ejection could be due to pulsational instabilities at the end of the evolution as suggested by \citet{Moriya2015a} just before a pair instability supernova. Of course, in that case we would expect to see also the products of the explosion, but in case the explosion were asymmetrical, as suggested by \citet{Gilmer2017a}, the only enrichment of the pre-stellar cloud could indeed be the matter lost through the pulsational matter ejection.
\section{Take-home message}
At low metallicity, massive stars are much more compact than at solar metallicity. Rotation is an important ingredient in their evolution, inducing a strong mixing. The nucleosynthesis is strongly affected by both the mixing and compactness, allowing the production of primary nitrogen and $s$-process elements. Evolving in a low-mass galaxy, they might enrich it only through slow matter ejection, either because they avoid a supernova explosion because of a too large CO core at the end of the evolution, or because the galaxy potential well is not able to keep the matter ejected at high speed during the explosion.

\end{document}